\documentclass{aa}
\usepackage{graphicx}
\begin{document}
   \title{Photometric amplitudes and phases of nonradial
          oscillation in rotating stars}

   \author{J. Daszy\'nska-Daszkiewicz$^{1,2}$, W. A. Dziembowski$^{2,3}$, A. A.
   Pamyatnykh$^{2,4}$, M-J. Goupil$^{5}$}

   \offprints{J. Daszy\'nska-Daszkiewicz, \email{daszynsk@camk.edu.pl}}

   \institute{{1} Astronomical Institute of the Wroc{\l}aw University,
   ul. Kopernika 11, 51-622 Wroc{\l}aw, Poland\\
   {2} Copernicus Astronomical Center, Bartycka 18, 00-716 Warsaw, Poland\\
   {3} Warsaw University Observatory, Al. Ujazdowskie 4, 00-478 Warsaw,
   Poland\\
   {4} Institute of Astronomy, Russian Academy of Sciences,
   Pyatnitskaya Str. 48, 109017 Moscow, Russia\\
   {5} DASGAL, UMR CNRS 8633, Observatoire de Paris-Meudon
    }

   \date{Received ...; accepted ...}

   \abstract{
Effects of rotational mode coupling on photometric parameters of
stellar oscillations are studied. At moderate rotation rates, a
strong coupling between modes of spherical harmonic degree,
$\ell$, differing by 2 and of the same azimuthal order, $m$, takes
place if the frequencies are close. This is a common situation
amongst main sequence pulsators. Numerical results for a sequence
of $\beta$ Cephei star models are reported for the two- and
three-mode couplings.\\
 One consequence of mode coupling is that modes of higher
degree should be considered in photometric mode identification.
Modes with nominal degree $\ell>2$ acquire substantial $\ell\le2$ components
and therefore are more likely to reach detectable amplitudes.
Coupled mode positions in the amplitude ratio -- phase
difference diagrams, based on multicolour photometry, become both
aspect- and $m$-dependent. Examples of the mode path in the
diagram with varying aspect are given. The diagrams remain a
useful tool for mode identification in rotating stars but the tool
must be used with care.
\\
   \keywords{stars: $\beta$ Cephei variables --
             stars: oscillation --
             stars: rotation
          }
}

   \titlerunning{Photometric amplitudes and phases of nonradial oscillation in rotating stars}
   \authorrunning{Daszy\'nska-Daszkiewicz et al.}
   \maketitle

\section{Introduction}
Mode identification, that is, determination of the radial order
and spherical harmonic, is an essential step in asteroseismology.
The task is not easy in the case of the oscillation frequency
spectra in $\beta$ Cep and $\delta$ Sct stars, which are most
often lacking equidistant patterns. The photometric diagnostic
diagrams, i.e., the amplitude ratio $vs.$ phase difference
dependencies in different passbands, are the most popular tools
for mode identification in pulsating stars. Following pioneering
works (Balona \& Stobie 1979, Stamford \& Watson 1981) these tools
have been applied mainly to $\beta$ Cep and $\delta$ Sct
variables.

Theoretical diagnostic diagrams are based on linear nonadiabatic
calculations of stellar oscillations and on models of static
plane-parallel atmospheres. In the early works an arbitrary
parametrization has been used instead of linear nonadiabatic
calculations. This approach has been followed even in some recent
studies (e.g. Garrido 2000). The nonadiabatic calculations were
first included explicitly by Cugier et al. (1994) and subsequently
by Balona \& Evers (1999), Cugier \& Daszy\'nska (2001), Balona et
al. (2001) and also by Townsend (2002), who applied them to SPB
stars.

Up to now, the amplitudes and phases, which we -- following Cugier
et al. (1994) -- call photometric nonadiabatic observables, were
calculated in the framework of linear nonadiabatic theory,
ignoring effects of rotation. However, amongst $\beta$ Cep and
$\delta$ Sct stars slow rotators, for which such an approximation
is adequate, are more an exception than a rule. Here we examine
effects of moderate rotation on theoretical diagnostic diagrams.
By moderate we mean so slow that perturbational treatment of
rotation is adequate. Specifically, we rely on the third order
formalism of Soufi et al. (1998).

The most important effect of rotation in the context of diagnostic
diagrams is coupling between close frequency modes of spherical
harmonic degree , $\ell$, differing by 2. The effect was discussed
in some detail by Soufi et al. (1998). The essential formulae are
recalled in the next section of this paper.

Numerical results presented later concern one selected sequence of
$\beta$ Cep models. On a qualitative level the results are
applicable to all stars of this type. Our choice of $\beta$
Cep stars is motivated not only by the abundance of the
observational data but also by the fact that we have credible
results from linear nonadiabatic calculations. This is not true
for $\delta$ Sct stars where there are serious uncertainties
related to the treatment of convection.

Properties of unstable modes in the selected sequence of models
are reviewed in Section 3. Also in this section we discuss the
occurence of near resonances between two and three modes as well
as certain consequences of mode coupling.

In Section 4 we discuss the visibility in various passbands of
modes described by a single spherical harmonic over a wide range of
$\ell$. Examples of diagnostic diagrams for coupled modes are
given in Section 5.

\section{Rotational mode coupling}

Chandrasekhar \& Lebovitz (1962) were the first to notice that
rotation, even at its slow rate, significantly influences mode
properties if there is a small frequency distance between modes
with the same azimuthal order, $m$, and spherical harmonic degree,
$\ell$, differing by 2. In this case, each of the coupled modes
must be represented by a certain superposition of spherical harmonics
of the two modes involved. The effect was invoked to explain
the nonradial character of pulsation in $\beta$ Cep stars. The effect
of mode coupling has been subsequently discussed in a number of
papers (e.g. Dziembowski \& Goode 1992, Soufi et al. 1998).

In the present paper we assume uniform rotation. Although our
calculations were carried out to cubic order in the ratio of
rotation to pulsation frequency, $\Omega/\omega$, for the purposes 
of this discussion and for the sake of simplicity, we will write  
formulae which are accurate only to $(\Omega/\omega)^2$.
This is enough to capture the effects we want to review here.

In general, there may be more than two coupled modes. Then the
displacement eigenfunction of an individual mode must be given as
a sum
$$\vec{\xi}=\sum_{k}a_{k} \vec{\xi}_{0k}, \eqno(1)$$
where $\vec{\xi}_{0k}$ stands for the eigenfunctions in the
spherical model. We choose the normalization in such a way that
the radial component of the vector $\vec{\xi}_{0k}$ is given by
$$\xi_{0,r}=RY_{l}^{m}\mathrm{e}^{-\mathrm{i}\omega t}. \eqno(2)$$
The orthogonality condition is expressed in the form
$$\int d^{3}\vec{x}\rho\vec{\xi}_{0j}\cdot\vec{\xi}_{0k}
=\delta_{jk}I_k, \eqno(3)$$
where $I_k$ is called  mode inertia. The eigenfuncions and
eigenvalues for coupled modes are calculated as a solution of the
matrix equation,
$$(\vec{\mathbf{B}}-\omega^2\vec{\mathbf{E}})\vec{a}=\vec{0},
\eqno(4)$$
where $\vec{\mathbf{E}}$ is a unit matrix and
$\vec{a}=(a_1, a_2, ...)$ (see e.g. Soufi et al. 1998). The
nondiagonal elements, $B_{jk}$, are of the order of $\Omega^2$ and
are non-zero for the modes satisfying conditions $\ell_j=\ell_k\pm
2$ and $m_j=m_k$. With our normalization these elements are not
symmetric. The following relation is fulfilled:
$$\frac{B_{kj}}{B_{jk}}=\frac{I_j}{I_k}. \eqno(5)$$
The diagonal elements are given by
$$B_{kk}=(\omega_{0k} +mC_k\Omega)^2 +{\cal O}(\Omega^2),
\eqno(6)$$
where $\omega_{0k}$ is the frequency in the spherically symmetric
model and $C_k$ is the Ledoux constant. The ${\cal O}(\Omega^2)$
term arises from quadratic effect of the Coriolis force and from
the centrifugal distortion. The latter effect dominates for
nonradial $p$-modes, as well as for g-modes of low orders.

Here we limit ourselves to couplings which involve up to three
modes. The solution in the case of two-mode coupling is
$$\omega^2_{\pm}=\frac{B_{11}+B_{22}\pm
\sqrt{(B_{11}-B_{22})^2+4B_C}}{2}, \eqno(7)$$
where $B_C=B_{12}B_{21}$, and
$$\left(\frac{a_1}{a_2}\right)_{\pm}=\frac{B_{12}}{B_{11}-\omega^2_{\pm}}.
\eqno(8)$$
The coupling strength, which is measured by $B_C$, depends on mode
properties. Coupling between acoustic modes is stronger than that
involving one or more gravity modes. This is so because the effect
of the centrifugal distortion is only important in the acoustic
cavity and it increases with the mode frequency.

 For an exact resonance, that is when $B_{11}=B_{22}$, we get
$$\left(\frac{a_1}{a_2}\right)_{\pm}=\mp{\rm sign}
(B_{12})\sqrt{\frac{I_2}{I_1}}, \eqno(9)$$
and, regardless of the coupling strength, we have strong mode
mixing. Then, the relative amplitudes are determined by mode
inertiae. Therefore, for instance, an $\ell=2$ mode trapped in the
interior thus having a large inertia may manifest itself as a
radial mode. A close doublet of "radial" modes may be detected
in this case.

Now we consider three-mode interaction, adopting $\ell_2=\ell_1+2$
and $\ell_3=\ell_2+2$. An interesting situation arises when the
closest resonance occurs between modes with $\ell$ differing by 4.
Since we have $B_{13}=B_{31}=0$, the interaction takes place
through the $\ell_2$ mode. Let us consider a simple limiting case
when $B_{11}=B_{33}$. Then, one solution is
$$\omega^2=B_{11}=B_{33},~~~ a_2=0,~~~
\frac{a_1}{a_3}=-\frac{B_{23}}{B_{21}}. \eqno(10)$$
The remaining two solutions are the same as in the case of two
mode coupling, except that $B_C=B_{12}B_{21}+B_{23}B_{32}$. The
highest degree mode, say $\ell=4$ or 5, which is normally
undetectable, acquires a large $\ell=0$ or, respectively, $\ell=1$
component. This means that some higher degree modes may become
detectable by means of ground-based photometry.

\section{Close frequencies in a sequence of ${\beta}$ Cephei star models}
  \begin{figure*}
  \centering
    \includegraphics[clip]{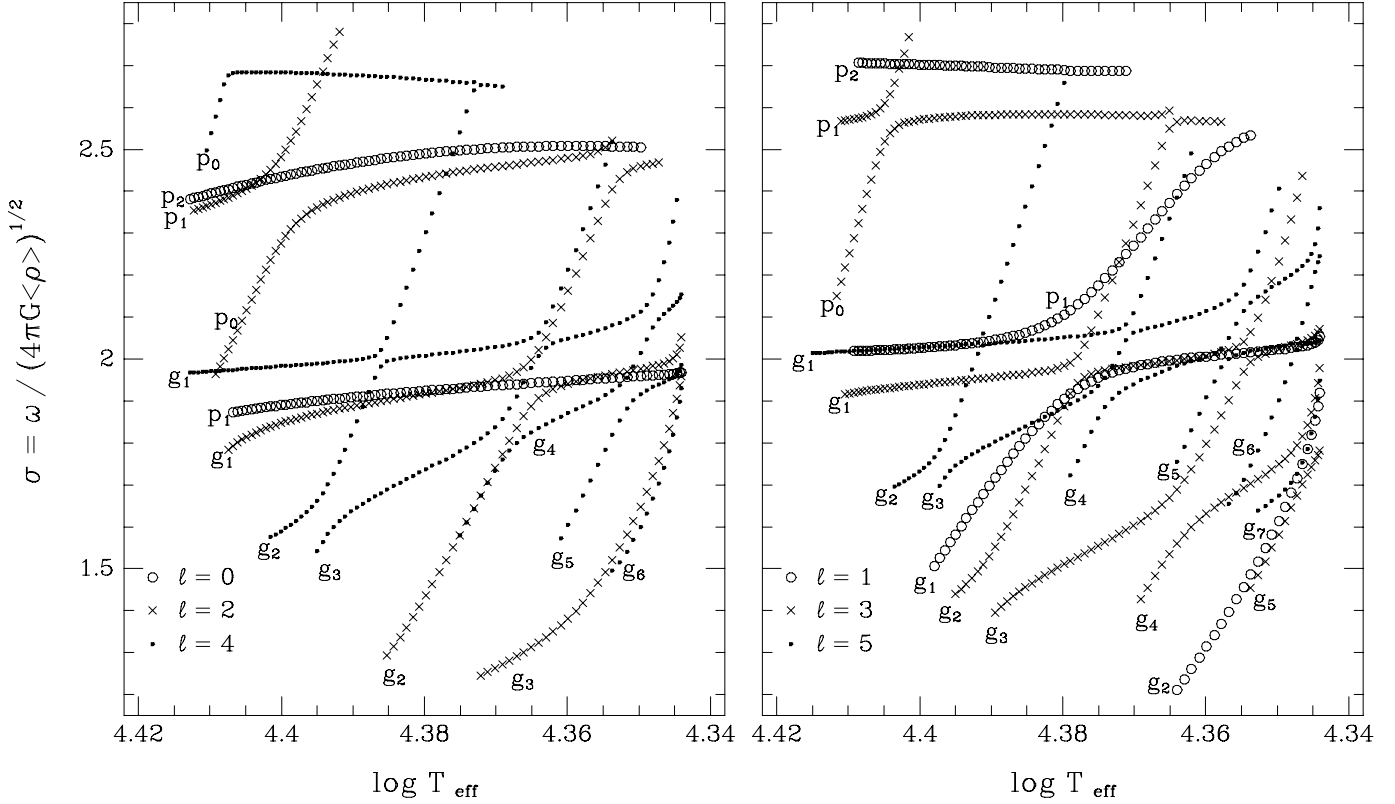}
             \caption{ Unstable low degree modes in a sequence of $\beta$
             Cephei star models with mass of 12 M$_\odot$. The
             models cover only the expansion phase of the main
             sequence evolution.
             In the left panel note that the $\ell=0,p_1$ and $\ell=2,g_1$
             modes have close frequencies in the extended
             range of $\log T_{\rm eff}$. At lower $\log T_{\rm eff}$
             the $\ell=2,g_2$ mode comes to a near resonance with the
             same radial mode. We see also the proximity of the nonradial
             $\ell=2$ and 4 modes at $\sigma\approx 1.6 - 1.8$.
             In the right panel, note a near
             resonance between the $\ell=1,g_1$ mode and, first, the $\ell=3,g_2$ and, later, the $g_3$.
             The proximity of three modes occurs only
             occasionally.}
        \label{aaaaaa}
         \end{figure*}

As an illustration we consider the evolutionary sequence of 12
$M_\odot$ star models in the $\beta$ Cep instability strip. We
adopted the standard chemical composition $(X=0.7,Z=0.02)$. Models
were calculated not allowing for convective overshooting from the
core. In these models we took into account the average effect
of centrifugal force, assumed uniform rotation with an equatorial
velocity of 100 km/s on the ZAMS, and imposed global angular
momentum conservation during evolution. The adopted value of rotational
velocity may be regarded typical for $\beta$ Cep stars. The coupling
strength is proportional to the square of the velocity, but the result
depends also on the frequency difference. Thus, we have no simple
scaling of the effect of rotational mode coupling.

Fig. 1 shows the frequency evolution in unstable low-degree
modes which are designated in accordance with the avoided-crossing
principle. The physical nature of the modes, that is, the relative
proportion of the contribution from acoustic and gravity
propagation zones to mode's energy, is reflected in the slope of
the $\sigma(\log T_{\rm eff})$ function. Slow and rapid rises
correspond to dominant acoustic and gravity mode characters,
respectively. The latter implies larger mode inertia.

We have to worry about mode coupling when the frequency distance
between the modes becomes comparable to the rotational frequency.
With our choice of rotation rate we have dimensionless rotational
frequency $\sigma_{\rm rot}\equiv\Omega/\sqrt{4\pi
G<\!\rho\!>}\approx0.1$. Thus, we can see that in many instances
the coupling may be significant.

Examples of coupled mode solutions are given in Table 1, where
only zonal modes ($m=0$) are included. The values of $\ell'$  were
assigned using again the avoided crossing principle, that is, the
ordering of modes in frequencies was kept unchanged. The relevant
frequency is $\nu_{NC}$, which corresponds to $B_{kk}$. In most
cases the value of $\ell'$ identifies the dominant component in
the surface amplitude. The exceptions occur for modes with
drastically different inertiae.

\begin{table*}
\begin{center}
\caption { Selected cases of rotational mode coupling. $\sigma_0$
is the dimensionless oscillation frequency in the spherical
models, $\nu_0=2\pi\omega_0$ is the corresponding cyclic frequency
in d$^{-1}$; $\nu_{NC}$ takes into account effects of rotation
except for mode coupling; $\Delta\nu_C$ is a frequency correction due to
coupling, whereas $\Delta\nu_T$ is a total frequency change
relative to the spherical model; $a_k$ are amplitudes of the
involved eigenmode components. }
\begin{tabular}{|ccccccrrcrrr|}
\hline Case & $\log T_{\rm eff}$ & coupling & $\sigma_0$ & $\nu_0$
& $\nu_{\rm_{NC}}$ & $\Delta\nu_{\rm_C}$ & $\Delta\nu_{\rm_T}$ &
coupled & $a_1$ & $a_2$ & $a_3$\\ &&&&&&&& modes &&&\\ \hline
 1  & 4.374 & $\ell_1=2$, $g_2$ & 1.611 & 3.836 & 3.849 & 0.002 & 0.014 & $\ell'=2$ & 0.88 &-0.47 &\\
    &       & $\ell_2=4$, $g_4$ & 1.610 & 3.834 & 3.846 &-0.002 & 0.011 & $\ell'=4$ & 0.62 & 0.78 &\\
\hline
 2  & 4.406 & $\ell_1=0$, $p_1$ & 1.875 & 6.438 & 6.457 & 0.002 & 0.022 & $\ell'=0$ & 1.00 &-0.07 &\\
    &       & $\ell_2=2$, $g_1$ & 1.801 & 6.182 & 6.215 &-0.002 & 0.031 & $\ell'=2$ & 0.14 & 0.99 &\\
\hline
 3  & 4.388 & $\ell_1=0$, $p_1$ & 1.913 & 5.265 & 5.282 & 0.013 & 0.030 & $\ell'=0$ & 0.91 &-0.42 &\\
    &       & $\ell_2=2$, $g_1$ & 1.894 & 5.214 & 5.248 &-0.013 & 0.021 & $\ell'=2$ & 0.51 & 0.86 &\\
\hline
 3A & 4.388 & $\ell_1=0$, $p_1$ & 1.913 & 5.265 & 5.282 & 0.011 & 0.028 & $\ell'=0$ & 0.92 & 0.36 &-0.15\\
    &       & $\ell_2=2$, $g_1$ & 1.894 & 5.214 & 5.248 &-0.014 & 0.020 & $\ell'=2$ & 0.50 &-0.86 & 0.05\\
    &       & $\ell_3=4$, $g_2$ & 1.921 & 5.288 & 5.303 & 0.003 & 0.019 & $\ell'=4$ & 0.59 & 0.51 & 0.63\\
\hline
 4  & 4.374 & $\ell_1=0$, $p_1$ & 1.932 & 4.600 & 4.615 &-0.015 & 0.000 & $\ell'=0$ & 0.85 & 0.53 &\\
    &       & $\ell_2=2$, $g_1$ & 1.934 & 4.605 & 4.636 & 0.015 & 0.046 & $\ell'=2$ & 0.58 &-0.82 &\\
\hline
 5  & 4.392 & $\ell_1=0$, $p_2$ & 2.462 & 7.096 & 7.110 & 0.008 & 0.022 & $\ell'=0$ & 0.99 &-0.16 &\\
    &       & $\ell_2=2$, $p_0$ & 2.385 & 6.877 & 6.913 &-0.008 & 0.028 & $\ell'=2$ & 0.24 & 0.97 &\\
\hline
 6  & 4.376 & $\ell_1=0$, $p_2$ & 2.496 & 6.064 & 6.076 & 0.015 & 0.027 & $\ell'=0$ & 0.95 &-0.30 &\\
    &       & $\ell_2=2$, $p_0$ & 2.445 & 5.939 & 5.976 &-0.015 & 0.022 & $\ell'=2$ & 0.39 & 0.92 &\\
\hline
 6A & 4.376 & $\ell_1=0$, $p_2$ & 2.496 & 6.064 & 6.076 & 0.015 & 0.027 & $\ell'=0$ & 0.95 & 0.30 & 0.00\\
    &       & $\ell_2=2$, $p_0$ & 2.445 & 5.939 & 5.976 &-0.015 & 0.022 & $\ell'=2$ & 0.39 &-0.92 & 0.00\\
    &       & $\ell_3=4$, $g_1$ & 2.491 & 6.050 & 6.058 & 0.000 & 0.009 & $\ell'=4$ & 0.17 &-0.06 &-0.98\\
\hline
 7  & 4.354 & $\ell_1=0$, $p_2$ & 2.508 & 4.875 & 4.886 &-0.019 &-0.009 & $\ell'=0$ & 0.91 & 0.41 &\\
    &       & $\ell_2=2$, $p_0$ & 2.521 & 4.899 & 4.925 & 0.019 & 0.046 & $\ell'=2$ & 0.59 &-0.81 &\\
\hline
 7A & 4.354 & $\ell_1=0$, $p_2$ & 2.508 & 4.875 & 4.886 &-0.020 &-0.009 & $\ell'=0$ & 0.91 &-0.42 & 0.00\\
    &       & $\ell_2=2$, $p_0$ & 2.521 & 4.899 & 4.925 & 0.020 & 0.046 & $\ell'=2$ & 0.59 & 0.81 & 0.01\\
    &       & $\ell_3=4$, $g_2$ & 2.518 & 4.895 & 4.902 &-0.000 & 0.007 & $\ell'=4$ & 0.71 & 0.27 &-0.64\\
\hline
\hline
 8  & 4.370 & $\ell_1=1$, $g_1$ & 1.981 & 4.532 & 4.566 &-0.023 & 0.011 & $\ell'=1$ & 0.73 & 0.68 &\\
    &       & $\ell_2=3$, $g_2$ & 1.983 & 4.537 & 4.571 & 0.023 & 0.057 & $\ell'=3$ & 0.67 &-0.75 &\\
\hline
 8A & 4.370 & $\ell_1=1$, $g_1$ & 1.981 & 4.532 & 4.566 &-0.018 & 0.016 & $\ell'=1$ & 0.79 &-0.60 &-0.12\\
    &       & $\ell_2=3$, $g_2$ & 1.983 & 4.537 & 4.571 & 0.027 & 0.061 & $\ell'=3$ & 0.61 & 0.79 & 0.10\\
    &       & $\ell_3=5$, $g_4$ & 1.940 & 4.439 & 4.456 &-0.009 & 0.009 & $\ell'=5$ & 0.09 &-0.42 & 0.90\\
\hline
 9  & 4.360 & $\ell_1=1$, $g_1$ & 2.006 & 4.144 & 4.180 & 0.022 & 0.058 & $\ell'=1$ & 0.75 &-0.66 &\\
    &       & $\ell_2=3$, $g_2$ & 2.008 & 4.148 & 4.178 &-0.022 & 0.007 & $\ell'=3$ & 0.72 & 0.70 &\\
\hline
 9A & 4.360 & $\ell_1=1$, $g_1$ & 2.006 & 4.144 & 4.180 & 0.000 & 0.036 & $\ell'=1$ & 0.87 & 0.00 &-0.49\\
    &       & $\ell_2=3$, $g_2$ & 2.008 & 4.148 & 4.178 &-0.044 &-0.014 & $\ell'=3$ & 0.38 &-0.71 & 0.59\\
    &       & $\ell_3=5$, $g_4$ & 2.006 & 4.143 & 4.180 & 0.044 & 0.081 & $\ell'=5$ & 0.39 & 0.69 & 0.61\\
\hline
 10 & 4.406 & $\ell_1=1$, $p_1$ & 2.021 & 6.938 & 6.988 & 0.004 & 0.055 & $\ell'=1$ & 1.00 &-0.08 &\\
    &       & $\ell_2=3$, $g_1 $ & 1.927 & 6.616 & 6.663 &-0.004 & 0.043 & $\ell'=3$ & 0.15 & 0.99 &\\
\hline
 10A& 4.406 & $\ell_1=1$, $p_1$ & 2.021 & 6.938 & 6.988 & 0.002 & 0.053 & $\ell'=1$ & 0.97 & 0.05 &-0.25\\
    &       & $\ell_2=3$, $g_1$ & 1.927 & 6.616 & 6.663 &-0.013 & 0.034 & $\ell'=3$ & 0.14 &-0.96 & 0.23\\
    &       & $\ell_3=5$, $g_1$ & 2.022 & 6.940 & 7.007 & 0.011 & 0.077 & $\ell'=5$ & 0.21 & 0.12 & 0.97\\
\hline
 11 & 4.372 & $\ell_1=1$, $p_1$ & 2.230 & 5.205 & 5.224 & 0.000 & 0.019 & $\ell'=1$ & 1.00 & 0.05 &\\
    &       & $\ell_2=3$, $g_1$ & 2.231 & 5.208 & 5.218 &-0.000 & 0.009 & $\ell'=3$ & 0.66 &-0.75 &\\
\hline
 12 & 4.380 & $\ell_1=1$, $p_2$ & 2.690 & 6.795 & 6.858 & 0.009 & 0.072 & $\ell'=1$ & 0.99 &-0.13 &\\
    &       & $\ell_2=3$, $p_0$ & 2.584 & 6.528 & 6.572 &-0.009 & 0.035 & $\ell'=3$ & 0.21 & 0.98 &\\
\hline
 12A& 4.380 & $\ell_1=1$, $p_2$ & 2.690 & 6.795 & 6.858 & 0.009 & 0.072 & $\ell'=1$ & 0.99 & 0.13 & 0.00\\
    &       & $\ell_2=3$, $p_0$ & 2.584 & 6.528 & 6.572 &-0.009 & 0.035 & $\ell'=3$ & 0.21 &-0.98 & 0.00\\
    &       & $\ell_3=5$, $g_2$ & 2.659 & 6.717 & 6.726 & 0.000 & 0.008 & $\ell'=5$ & 0.08 &-0.17 &-0.98\\
\hline

\end{tabular}
\end{center}
\end{table*}

The cases of exceptionally close two-mode resonances are 1, 3, 4,
7, 8, 9 and 11. There are differences between these cases, which
reflect not only the differences in the proximity of the resonance
but also in the nature of the coupled modes. As we have already
discussed in section 2, the coupling involving modes of gravity
character is weaker than that involving only modes of the acoustic
character. Weak coupling is reflected in the low values of
$\Delta\nu_C$ in cases 1 and 11. However, the mixing of the modes
is usually strong, except for $\ell'=1$ mode in case 11, which
remains essentially pure $\ell=1$. This has to do with the
disparity of inertiae for the coupled modes. In cases 3 and 4 both
modes have $p$-mode character and the correction due to coupling
is larger. Still, the frequency shift of $\sim0.015$ d$^{-1}$ is
significant only for data obtained with observations lasting
longer than 2 months. In general, we find rather small rotation
induced frequency changes, even in those cases when the coupling
is large.

In some of the listed cases, like 2, 10 and 12, the coupling
appears quite weak. Nonetheless, as we will see later in section
5, the diagnostic diagrams may be significantly perturbed.

Among the three-mode couplings, case 9A most closely corresponds
to the situation described in section 2. With the accuracy
adopted, the modes $\ell=1$ and 5 are in the exact resonance and
the $\ell=3$ mode is not far away. Also in cases 3A and 7A, the
three-mode coupling is important. In the latter case we see that
$\ell'=4$ mode acquires a large $\ell=0$ component. In case 10A
the frequency distance between $\ell=1$ and 5 is small, but
because the distance to $\ell=3$ is relatively large, the coupling
is essentially inconsequential. Also in case 6A, the coupling to
the third mode is without significant consequences. Here the
reason is the gravity character of $\ell=4$ mode.

Coming to tesseral modes ($m\ne0$), we first note that the
coupling occurs if $m=\pm1$ and, for even $\ell$ modes, also if
$m=\pm2$. In general, coupling conditions between prograde and
retrograde mode pairs are not the same. The asymmetry is due to
the difference in the Ledoux constant for the coupled modes. The
only systematic difference relative to the $m=0$ modes is a
somewhat weaker coupling due to smaller value of the angular
integrals entering expressions for $B_{jk}$.

The near degeneracy of rotationally coupled modes occurs also in
wide ranges of $\delta$ Sct star models. For high order $p$-modes
the effect is systematic, as implied by the asymptotic relation
$\omega_{n,\ell+2}\approx\omega_{n-1,\ell}$. Since the coupling
strength increases with frequency, in the high order $p$-mode
pulsators the effect begins to be significant at lower rotation
rates than in $\beta$ Cep stars. Frequency changes caused by
rotational mode coupling in $\delta$ Sct stars and in solar type
pulsators have been discussed by Goupil et al. (2000) and by
Dziembowski \& Goupil (1998), respectively.

\section{Photometric diagnostic diagrams for single spherical harmonic modes}

Our approach follows that of Cugier et al. (1994), except for
correcting an error in the expression for perturbed gravity, which
fortunately has negligible consequences (Cugier \& Daszy\'nska,
2001). Here we outline the main steps. In our construction of
theoretical diagnostic diagrams we rely on the linear nonadiabatic
description of oscillations. Normalization of eigenfunctions is
fixed in Eq. 2. With that, the local displacement of the
photosphere in the stellar reference frame may be written as
follows
$$\delta r(R,\theta,\varphi)=\varepsilon R {\rm Re}\{ Y_\ell^m
{\rm e}^{-{\rm i}\omega t}\}, \eqno(11)$$
where $\varepsilon$ is a small quantity, which is determined by
data. This equation implies that the phase zero corresponds to the
maximum of displacement. The use of $-{\rm i}\omega t$ in the
exponential time dependence, which is now the most common choice,
instead of $+{\rm i}\omega t$ used by Cugier et al. (1994),
results in the opposite signs for the phase differences between
calculated observables. For a comparison of calculated and
measured values it is important to check the sign convention
adopted in the data analysis. With the sign convention adopted in
the present work, the phase difference $\phi_a-\phi_b>0$ means
that the maximum of $a$ occurs after the maximum of $b$.

The corresponding changes of the effective temperature and gravity
during the pulsation cycle are given by
$$\frac{ \delta T_{\rm eff} } { T^0_{\rm eff} }= \varepsilon
\frac14 {\rm Re}\{ f Y_\ell^m {\rm e} ^{-{\rm i} \omega t} \},
\eqno(12)$$
and
$$\frac{\delta g_{\rm eff}}{g_{\rm eff}^0} = - \left( 2 +
\frac{ 3\omega^2}{4\pi G<\!\rho\!>} \right) \frac{\delta r}{R},
\eqno(13) $$
respectively, where $f$ is the complex quantity determined from
linear nonadiabatic calculations, $<\!\rho\!>$ is the mean density
of the star and $G$ is the gravitational constant.

Further, we assume static plane-parallel atmosphere, and we write
the monochromatic flux variation caused by the oscillation mode
with $\ell$ degree in the following form:
$$\frac{\Delta{\cal F}_{\lambda}}{{\cal F}_{\lambda}^0} =
\varepsilon Y_{\ell}^m(i,0) b_{\ell}^{\lambda} {\rm Re} \{ [
D_{1,\ell}^{\lambda} +D_{2,\ell} +D_{3,\ell}^{\lambda}] {\rm
e}^{-{\rm i}\omega t} \}, \eqno(14)$$
where
$$D_{1,\ell}^{\lambda} = \frac14  f \frac{\partial \log ( {\cal
F}_\lambda |b_{\ell}^{\lambda}| ) } {\partial\log T_{\rm{eff}}} ,
\eqno(15a)$$
$$D_{2,\ell} = (2+\ell )(1-\ell ), \eqno(15b)$$
$$D_{3,\ell}^{\lambda}= -\left( \frac{3\omega^2}{4\pi G<\!\rho\!>}
+ 2 \right) \frac{\partial \log ( {\cal F}_\lambda
|b_{\ell}^{\lambda}| ) }{\partial\log g_{\rm eff}^0} \eqno(15c)$$
and
$$b_{\ell}^{\lambda}=\int_0^1 h_\lambda^0(\mu) \mu P_{\ell}(\mu)
d\mu, \eqno(16)$$
 In this formula, the $D_{1,\ell}^\lambda$ term describes the temperature
effects, whereas the influence of the gravity changes is contained
in the $D_{3,\ell}^\lambda$ term. Both include the perturbation of
the limb-darkening. In our calculation we relied on Claret's
(2000) analytical fit to Kurucz's (1998) tabular data. The
$\ell$-dependence of $D_1$ and $D_3$ arises from the nonlinearity
of the adopted limb-darkening law. The $D_{2,\ell}$ term stands
for the geometrical effects. The effect of orientation  with respect
to the observer is described by the spherical harmonic $Y_\ell^m(i,0)$.

  \begin{figure}
  \centering
    \includegraphics[width=88mm,clip]{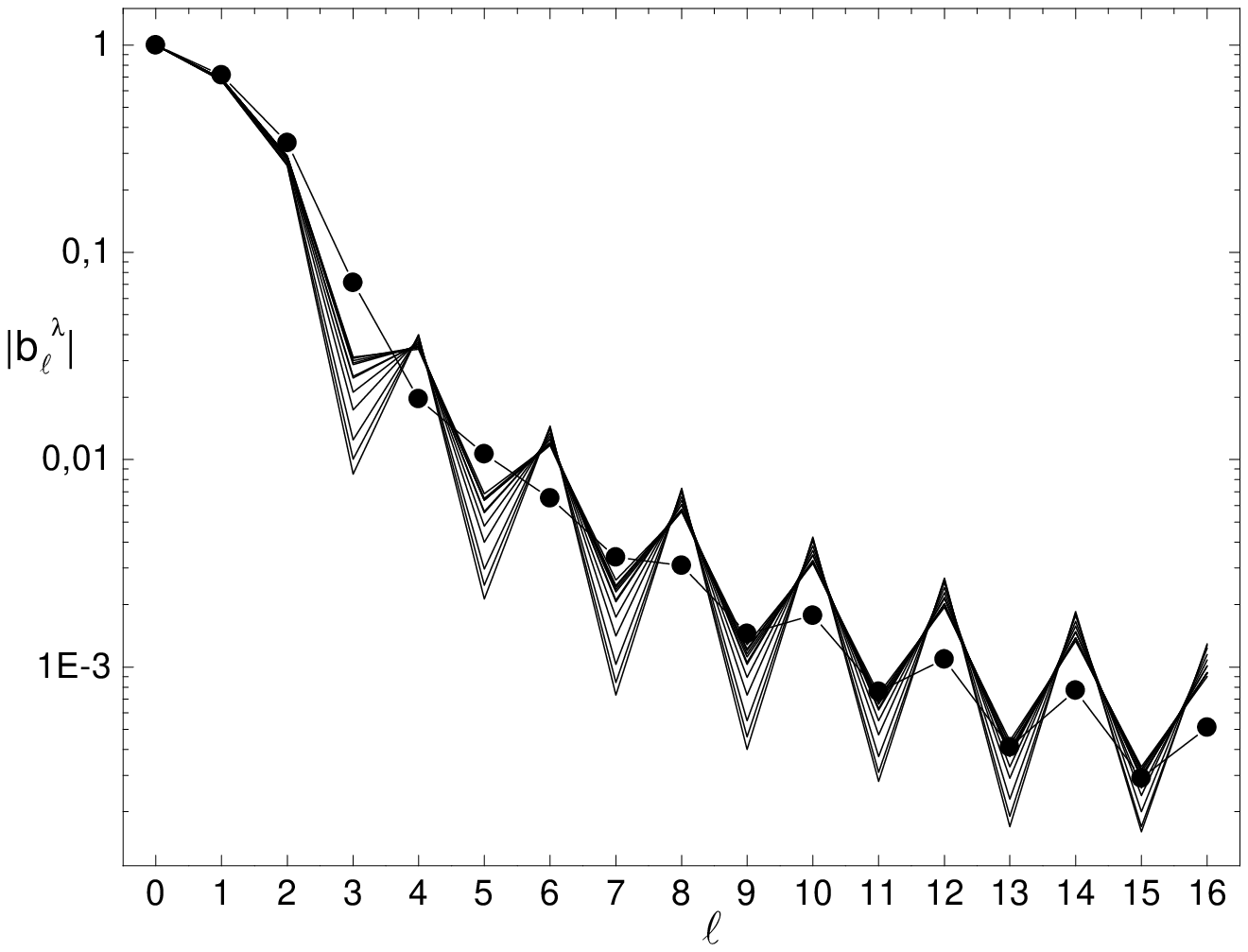}
             \caption{ The absolute values of the disc averaging
             factor plotted for bolometric and for twelve passbands
             as a function of $\ell$. The passbands include
             $u,v,b,y$ and $U,B,V,R,I,J,H,K$ filters.
              }
        \label{aaaaaa}
         \end{figure}

 In Fig. 2 we show the dependence of the disc averaging factor,
$b_{\ell}^{\lambda}$, on mode degree, $\ell$, for different
passbands. For comparison, the bolometric values are also shown.
These are similar to those calculated by Dziembowski (1977) with
the Eddington's limb darkening law. In contrast, large differences
are seen for individual passbands. Stronger averaging of
odd-$\ell$ modes gives rise to a pronounced oscillatory dependence
on $\ell$, which is superimposed on the decreasing trend. The
amplitude is larger than for the bolometric flux due to the
flatter limb-darkening law. Note that in the case of uniform
brightness disc, the odd-$\ell$ modes beginning with $\ell=3$ are
completely averaged out.

The decreasing trend does not properly reflect the $\ell$
dependence of the expected amplitudes. This would be true only if
the temperature term dominates. For modes of low radial orders the
geometrical term $D_2$ becomes dominant around $\ell=6$. Then
owing to the $(\ell+2)(\ell-1)$ factor the decline with $\ell$ is
slower than implied by Fig.2.

For calculating diagnostic diagrams it is convenient to use the
amplitude of the monochromatic flux variation in the following
complex form
$$A_\lambda(i) = \varepsilon Y_\ell^m(i,0) b_\ell^\lambda (
D_{1,\ell}^{\lambda} +D_{2,\ell} +D_{3,\ell}^{\lambda}), \eqno(17)
$$
which is equivalent to Eq.14. From this expression we can directly
obtain the amplitude ratio and phase difference in any selected
pair of passbands. Our choice is $u$ and $y$ Str\"omgren filters.
For modes described by a single $Y_\ell^m$, these photometric
observables are independent of the aspect angle and azimuthal
order, $m$.

  \begin{figure}
  \centering
    \includegraphics[width=88mm,clip]{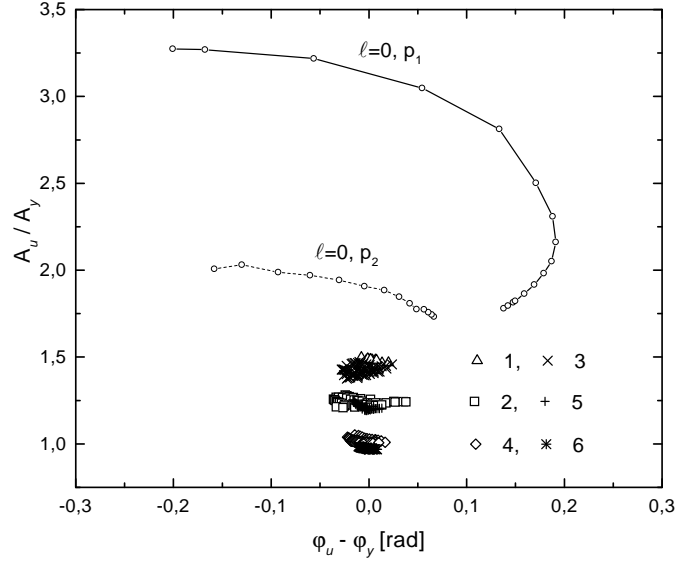}
             \caption{ Theoretical diagnostic diagram with the
              Str\"omgren passbands $u$ and $y$ for unstable low degree modes in the
              $\beta$ Cep star models with $M=12M_\odot$. In the
              sequences for the $\ell=0$ modes the most negative
              values of the phase difference correspond to the
              coolest models.
              }
        \label{aaaaaa}
         \end{figure}

The diagnostic diagram for unstable modes in our model sequence is
presented in Fig. 3. Modes up to $\ell=6$ are included. We can see
that, in contrast to the radial modes, the nonradial modes are
crowded in small areas. Two aspects of this plot call for an
explanation. Firstly, why it is that the amplitude ratios and phase
differences of radial modes are particularly dependent on
frequency and model temperature?
Secondly, what is the origin of the grouping of nonradial modes?

The main reason for this distinctive behavior of radial modes is a
comparable but opposite contribution of the temperature and the
geometrical terms to luminosity variation. The sum of the real part of
$D_1^\lambda$ and $D_2$ varies significantly with temperature and
it is strongly frequency dependent. For $\ell=1$, we have $D_2=0$
and the only reason for the spread of the mode positions is a weak
dependence of $b_\ell^\lambda$ on the passbands. The role of the
geometrical term for $\ell=2$ modes is comparable to that of
$D_1^\lambda$. However, the two terms add, which results in much
smaller spread of their positions. The near overlap of $\ell=1$
and 3 modes is coincidencial. At still higher $\ell$, when
geometrical effect dominates, the even and odd degree modes are
gathered in two small separate domains, determined solely by the
$b_\ell^\lambda$ factor (see Fig. 2). The phase difference is
always close to zero owing to the small role of the only complex
term, $D_1^\lambda$.

\section{Photometric diagrams for coupled modes}

The complex amplitude of the monochromatic flux variation for a
coupled mode may be expressed in the form
$${\mathcal A}_\lambda(i)= \sum_k a_k A_{\lambda,k} (i),
\eqno(18)$$
where $a_k$ are solutions of Eq. 4 and $A_\lambda(i)$ is given in
Eq.17. The quantities $a_k$ describe contributions of the
$\ell_k$-modes to the coupled mode $\ell'$. Selected values of
$a_k$ are given in Table 1. The quantity $f$ appearing in Eq. 17
is only weakly dependent on $\ell$. Its relatively strong
dependence on frequency is irrelevant in view of very small
frequency spread of the mode involved. In contrast to the
amplitude ratio and phase difference for the pure modes, these
quantities now depend on the inclination angle, $i$, and the
azimuthal order, $m$. We will show here few representative
examples of the photometric diagnostic diagrams in the case of two
and three mode coupling and we will discuss the consequence for
the mode identification.

Naively, one could expect that the coupled mode appears in the
diagnostic diagrams between positions corresponding to its
components associated with individual spherical harmonics. We will
see that this is not true.

As our first illustrative example, we chose the case when the
$\ell=0$ and 2 modes are in close resonance (case 4 in Table 1).
In Fig. 4 we show how positions of the coupled modes move in the
diagnostic diagram with varying aspects.
  \begin{figure}
  \centering
    \includegraphics[width=88mm,clip]{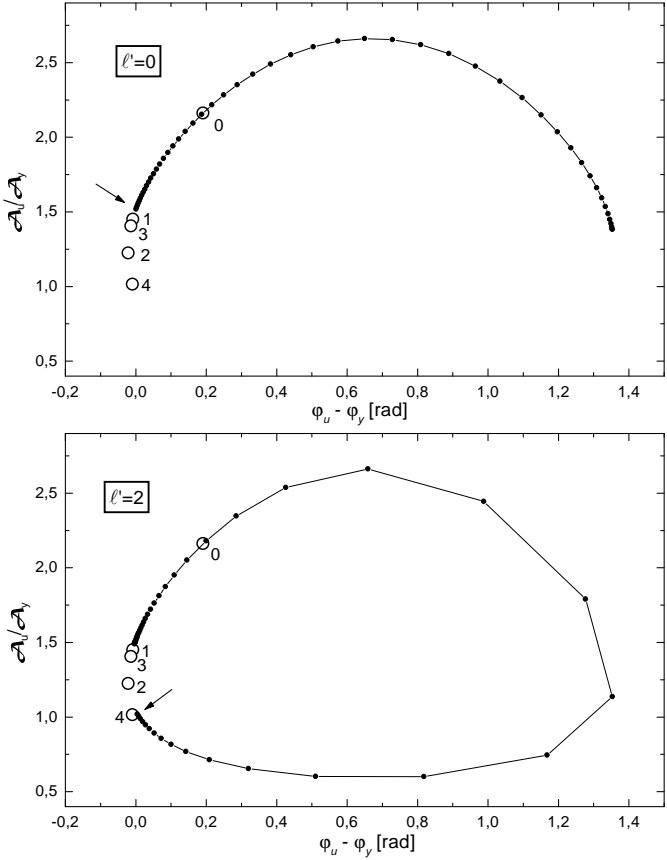}
             \caption{ The diagnostic diagrams with
             Str\"omgren photometry passbands $u$ and $y$,
             showing positions for coupled $\ell=0,p_1$ and
             $\ell=2,g_1$ modes (case 4 in Table 1) in the model at $\log T_{\rm eff}=4.3741$
             (see Fig. 1). This model corresponds to the
             smallest frequency distance between these two modes.
             The upper panel refers to the solution dominated by the $\ell=0$
             component, while the lower one refers to the solution
             dominated by $\ell=2$ component.
             Arrows correspond to observations from the polar direction.
             Spacing between consecutive dots is 0.02 in $\cos i$.
             Note that the density of dots reflects probability of
             observing the modes in various parts of that diagram.
             Circles indicate positions of pure $\ell=0,1,2$ modes.
              }
        \label{aaaaaa}
         \end{figure}
  \begin{figure}
  \centering
    \includegraphics[width=88mm,clip]{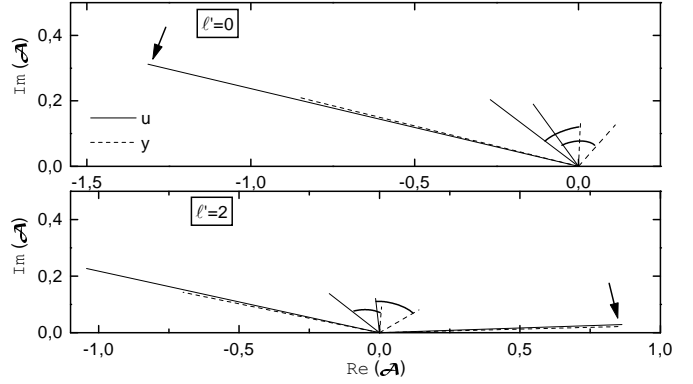}
             \caption{ Complex amplitudes in Str\"omgren passbands $u,~y$ for modes used in Fig.4.
             The three pairs of vectors
             shown in the upper panel correspond to $\cos i=1.0,0.32$ and 0.0.
             The four pairs of vectors in the lower panel correspond to $\cos i=0.0,0.66,0.72$
             and 1.0. The arrows mark amplitudes seen from the polar direction.}
        \label{aaaaaa}
         \end{figure}
A comparison of plots in the upper and lower panels reveals that
the movement is strongly mode-dependent. Let us focus first on the
case of mode dominated by the radial component, shown in the upper
panel. Due to close resonance, contamination with the $\ell=2$
component is significant. The amplitudes of the $\ell=0$ and 2
components are 0.85 and 0.53, respectively. An observer from the
polar direction will identify this mode as $\ell=1$. This is so,
because the main role of the $\ell=2$ component is a cancellation
of the geometrical term in the expression for the flux variation
(see. Eq. 14). When the observer moves away toward the equatorial
direction, he will see the mode as a pure $\ell=0$, at the angle
corresponding to the node of $Y_2^0(i,0)$, and subsequently he
will get mode characteristics which do not correspond to any of
single-$\ell$ modes. Note that such an appearance has a
significant probability, as measured by the density of dots.
  \begin{figure}
  \centering
    \includegraphics[width=88mm,clip]{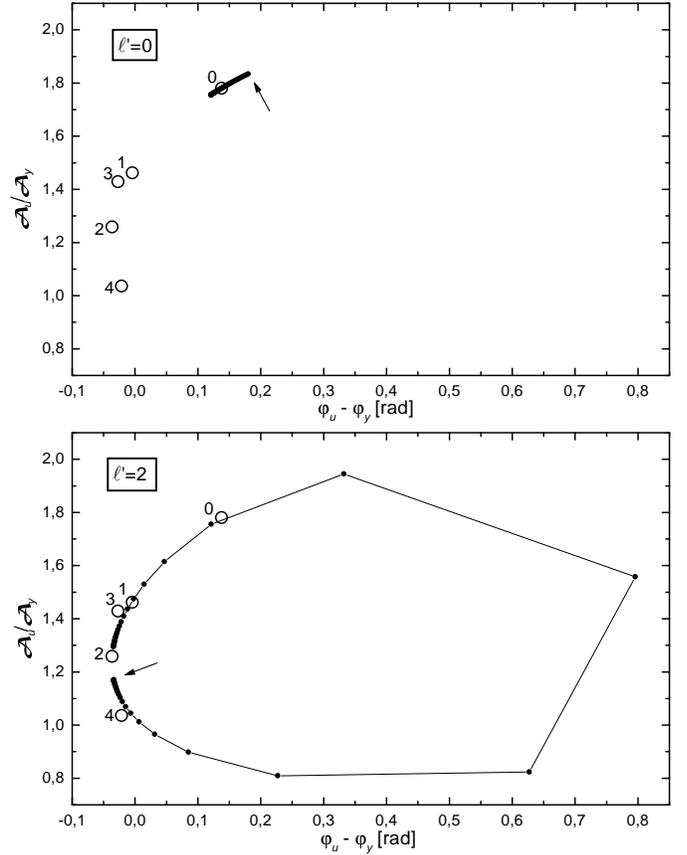}
             \caption{ The same as in Fig.4 but the coupling between modes
              $\ell=0$ and 2 is much weaker (case 2 in Table 1).}
        \label{aaaaaa}
         \end{figure}

Fig. 5 helps us to understand the cause of the described
behaviour. Here we show complex, arbitrarily normalized,
amplitudes in the two selected passbands. To make the amplitudes
realistic one should divide them by a factor between 10 to 100. We
see an increasing phase difference between two passbands when we move
from polar to equatorial direction. This is predominantly due to a
decrease of the real part of the complex amplitude in $y$ passband.
Why does the argument for the $y$ passband change much more
rapidly with aspect than that for $u$ passband? This is due to a
smaller absolute value of $D_1$, hence the greater role of $D_2$.
Only in this colour does ${\rm Re}({\cal A})$ pass through zero.
This happens at $\cos i=0.32$. Note, however, that when the phase
difference is large, the pulsation amplitude is rather small. So
that the situation may not be easily observable.

The very different pattern of the $\ell'=2$ mode behaviour, shown in
lower panel of Fig.4, is due to the fact that now the real part of
the amplitude in both colours changes sign. This is caused by the
greater contribution from the $\ell=2$ component. Here the
equatorial position of the mode is not so distant from the polar
one. Surprisingly, the mode is often seen as either $\ell=1$ or
$\ell>3$.

  \begin{figure}
  \centering
    \includegraphics[width=88mm,clip]{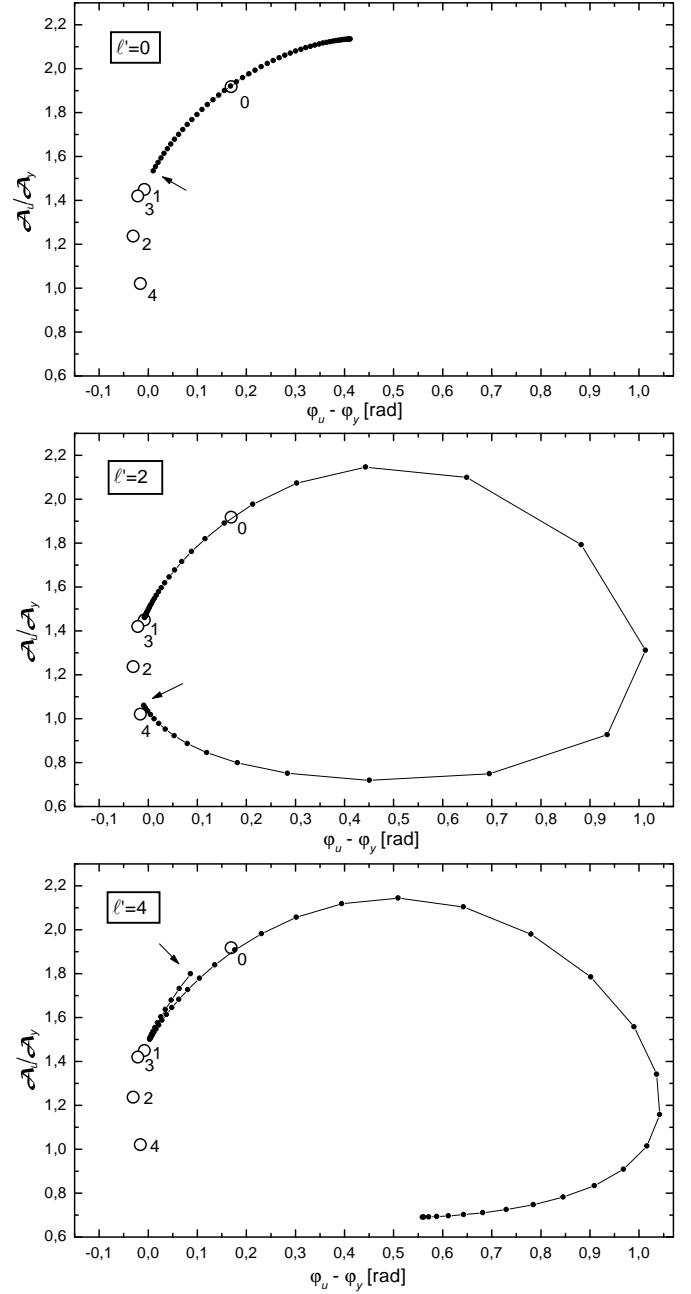}
             \caption{An example of the diagnostic diagrams for the three mode coupling
             (case 3A).}
        \label{aaaaaa}
         \end{figure}

Fig. 6 illustrates case 2, when as we have noted the coupling is
rather weak. Whereas for the $\ell'=0$ root the effect of the
coupling is marginal, for $\ell'=2$ it is quite significant. In
spite of the fact that the mode is dominated by $\ell=2$
component, it has a fairly high chance to be identified as
$\ell=1$ or $\ell>3$.

The cases of $\ell=1$ and 3 coupling never lead to a large
departure of the $\ell'=1$ modes from the $\ell=1$ position. Since
mode $\ell'=3$ acquires a substantial $\ell=1$ component, it is
mostly seen through this component and hence would be identified
as $\ell=1$. It is interesting that many different modes in
rotating stars may be identified as $\ell=1$ by means of the
diagnostic diagram method, if effects of rotation are ignored.

Fig. 7 illustrates an interesting case (case 3A in Table 1) of the
three-mode coupling. The $\ell'=0$ mode is most often seen close
to the $\ell=0$ position, but there is also a fair probability to
see this mode close the $\ell=1$ position. The path for the
$\ell'=2$ mode is very similar to that shown in the lower panel of
Fig.5. The $\ell'=4$ mode is most often seen between $\ell=0$ and
$\ell=1$ positions. However, seen from the equatorial direction it
is quite far from any position of a single $\ell$ mode.

For other modes of the same multiplets the coupling effects are
less spectacular. In all the cases from $m= -2$ to $m=+2$, the
$\ell'=2$ modes hardly move from the $\ell=2$ position. As
expected, the effect for the $\ell'=4$ mode is more significant.
The mode stays predominantly between the $\ell=1$ and $\ell=4$
positions.

Fig. 8 shows how mode-coupling of the multiplets would be identified 
if rotation is neglected. In this figure we plotted the observed amplitudes
assuming that the intrinsic mode amplitudes are the same. Note that the
indicated amplitudes should be divided by a factor between 10 and
100 to correspond to reality.
Identifications were based on the diagram in Fig.3.

  \begin{figure}
  \centering
    \includegraphics[width=88mm,clip]{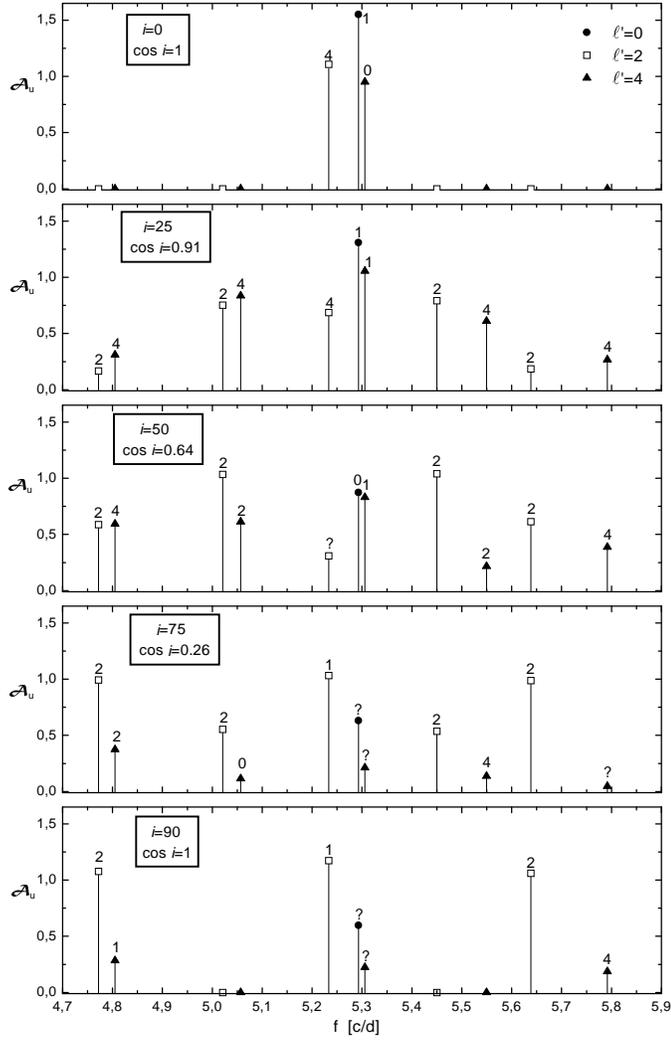}
             \caption{ Multiplets of coupled $\ell'=0,2,4$ modes for case 3A.
             Amplitudes in the $u$ band were calculated assuming common
             normalization, $\varepsilon=1$ in Eq.17. Numbers above the symbols give
             most likely identification upon neglecting effects of rotation. Lack of numbers
             means that mode can not be identified with any $\ell$.}
        \label{aaaaaa}
         \end{figure}

\section{Conclusions}

We have seen that even at moderate rotation, mode coupling leads to
complications in the diagnostic diagrams used for photometric mode
identification. Modes of degrees $\ell>2$, which are often
ignored in identification of peaks in oscillation spectra, may
acquire substantial low $\ell$ components, and are more easily detected.
Unlike modes described by a single spherical
harmonic, the positions of coupled modes in the diagnostic
diagrams depend on the aspect and on the azimuthal order. The
positions may be quite confusing, for instance, a mode composed of
the $\ell=0$ and $\ell=2$ components may appear for a range of
aspect angles at the $\ell=1$ position.

All that is not good news for the photometric mode identification
procedure. The problems are not confined to a few cases, but
occur at the typical rotation rates
encountered in $\beta$ Cep and $\delta$ Sct stars. Close
frequencies of rotationally coupled modes occur over wide ranges
of the instability strips. The implication is that we must be
careful in using the diagnostic diagrams for inferring the $\ell$
values. The diagrams remain useful. After all, they do provide
observational constrains on stars and their oscillations. However,
mode identification may be done only simultaneously with
determination of stellar parameters and inclination of rotational
axis.

\acknowledgements{J. D.-D. gratefully acknowledges hospitality of
the Copernicus Astronomical Center staff during her postdoctoral
stay. The work was supported by KBN grant No. 5 P03D 012 20.}


\begin{thebibliography}{}

   \bibitem[1979]{balona79} Balona, L. A., Stobie, R. S., 1979, MNRAS 189, 649

   \bibitem[1999]{balona99} Balona, L. A., Evers, E. A., 1999, MNRAS 302, 349

   \bibitem[2001]{balona01} Balona, L. A., Bartlett, B., Caldwell, J. A. R., et al., 2001,
    MNRAS 321, 239

   \bibitem[2000]{claret00} Claret, A., 2000, A\&A 363, 1081

   \bibitem[1962]{chandra62} Chandrasekhar, S., Lebovitz, N. R., 1962, ApJ 136,
    1105

   \bibitem[2001]{cugier01} Cugier, H., Daszy\'nska, J., 2001, A\&A 377, 113

   \bibitem[1994]{cugier94} Cugier, H., Dziembowski, W. A., Pamyatnykh, A. A. 1994,
         A\&A 291, 143

   \bibitem[1977]{dziemb77} Dziembowski, W. A., 1977, Acta Astron. 27, 203

   \bibitem[1992]{dziemb92} Dziembowski, W. A., Goode, P. R., 1992, ApJ 394, 670

   \bibitem[1998]{dziemb98} Dziembowski, W. A., Goupil, M.-J., 1998,
    in {\it Workshop on Science with Small Space Telescopes}, eds. T.R. Bedding
    and H. Kjeldsen (University of Aarhus), p. 69.

   \bibitem[2000]{garrido00} Garrido, R.,2000,
   in {\it Delta Scuti and Related Stars}, Proc. of the 6th Vienna Workshop in
   Astrophysics,  eds. M. Breger and M. H. Montgomery, ASP Conf. Ser., Vol. 210, p.67

   \bibitem[2000]{goupil00} Goupil, M.-J., Dziembowski, W. A., Pamyatnykh, A. A., Talon, S.,2000,
   in {\it Delta Scuti and Related Stars}, Proc. of the 6th Vienna Workshop in
   Astrophysics,  eds. M. Breger and M. H. Montgomery, ASP Conf. Ser., Vol. 210, p.267

   \bibitem[1998]{kurucz96} Kurucz, R. L. 1998, CD-ROM No. 13 and 19

   \bibitem[1998]{soufi98} Soufi, F., Goupil, M.-J., Dziembowski, W. A., 1998,
         A\&A, 334, 911

   \bibitem[2002]{towsend02} Townsend R. D., 2002, MNRAS 330, 855

   \bibitem[1988]{watson88} Watson R. D., 1988, Ap\&SS 140, 255

\end{thebibliography}
\end{document}